\documentclass[12pt,a4paper]{article}
\usepackage{amssymb}
\newcounter{theorem}
\newtheorem{theorem}{Theorem}[section]   
\newcounter{eg}
\newtheorem{eg}{Example}[section]   
\def\beg{\begin{eg}\rm}
\def\eeg{\hfill\sq\end{eg}}
\def\t#1{\tilde #1}
\def\ad{\mbox{ad}\,}
\def\b#1{{\mathbb #1}}

\def\c#1{{\cal #1}}

\def\Dirac{{\raise0.09em\hbox{/}}\kern-0.69em D}

\def\kbar{{\mathchar'26\mkern-9muk}}  

\def\sq{\hbox{\rlap{$\sqcap$}$\sqcup$}}

\def\p{\partial}

\def\tr{\mbox{Tr}}

\def\wm{\mathbin{*}}
\def\hwm{\mathbin{\hat *}}
\def\k{\kern-.1em\mathbin{,}\kern-.1em}
\def\hk{\kern.12em\raise-1em\hbox{$\hat{\raise1em\hbox{,}}$}\kern.12em}
\newcommand{\initiate}{\setcounter{equation}{0}}
\def\cn#1{\cite{#1}}

\begin{document}

\title{Fuzzy-{$pp$} Waves}

\author{J. Madore$\strut^{1,2}$ \ M. Maceda$\strut^{1}$ \
        D.~C. Robinson$\strut^{3}$\\[10pt]
        $\strut^{1}$Laboratoire de Physique Th\'eorique\\
        Universit\'e de Paris-Sud, B\^atiment 211, F-91405 Orsay
\and    $\strut^{2}$Max-Planck-Institut f\"ur Physik\\
        F\"ohringer Ring 6, D-80805 M\"unchen\\[5pt]
\and    $\strut^{3}$Dept. of Mathematics\\
        King's College, Strand, UK-WC2~R2LS London}

\date{}

\maketitle

\abstract

We present a noncommutative version of a plane-wave solution to the
gravitational field equations. We start with a given classical
solution, admittedly rather simple, and construct an algebra and a
differential calculus which supports the metric. In the particular
solution presented as an example the 1-forms do not anticommute, to a
degree which depends on the amplitude of the deviation of the metric
from the standard Minkowski metric.

\vfill

\medskip
\eject

\parskip 4pt plus2pt minus2pt

\initiate
\section{Motivation}

A noncommutative generalization of a classical solution to Einstein's
equation would require first of all a noncommutative version of a
smooth manifold. Although noncommutative geometry has its roots in
several attempts by physicists to overcome the problems of ultraviolet
divergences in field theory it is to A.~Connes that is due the first
construction~\cite{Con94} of a smooth structure. The version of this
which we shall use here~\cite{Mad00c} is a slight, even `smoother',
modification which permits the use of a metric of Minkowski signature.
This metric has also a locality property which is most easily
expressible in terms of a `frame' in the sense of Cartan. A smooth
manifold $V$ with a metric can be defined as a subspace of an
euclidean space $\b{R}^m$ of sufficiently high dimension $m$ and the
embedding can be chosen so that the induced metric coincides with that
of the manifold. If the manifold is of dimension $n$ then generically
in this case $m$ must satisfy $m \geq n(n-1)/2$. The embedding is
equivalent to a projection 
$$
\c{C}(\b{R}^m) \rightarrow \c{C}(V)
$$
of the algebra of functions $\c{C}(\b{R}^m)$ onto the algebra
$\c{C}(V)$ of smooth functions on $V$ and the latter can be formally
defined in terms of generators and relations.  This is convenient for
the noncommutative generalizations. 

Classical metrics have often been
defined~\cite{LidRomTavRip97,PavTap01} using embeddings and although
it is possible in the present case to avoid this
construction~\cite{DesPirRob76} it will perhaps in general
be convenient.  Other metrics with curvature have been
proposed~\cite{Mad89c,AscCas93} in the past but none
has been found to be Einstein. The method we use is for the moment
limited to what could be considered noncommutative analogues of
parallelizable manifolds. Another popular approach does not have this
limitation but is restricted to metrics of euclidean signature; we
refer to two recent books~\cite{Lan97,FigGraVar00} for an exhaustive
description and reference to the original literature.

Within the general framework which we here consider, the principal
difference between the commutative and noncommutative cases lies in
the spectrum of the operators which we use to generate the
noncommutative algebra which replaces the algebra of functions. This
in turn depends not only on the structure of this algebra as an abstract
algebra but on the representation of it which we choose to
consider. The ultimate aim is the construction of a noncommutative
generalization of the theory of general relativity which will remain
``smooth', but become essentially noncommutative, in regions where the
commutative limit would be singular. An explicit example has been
constructed~\cite{MacMadZou02} recently based on the Kasner metric.

\initiate
\section{The general formalism}

For the purpose of the present exposition the expression
`noncommutative space-time' will designate a $*$-algebra $\c{A}$ with
trivial center and $n(=4)$ hermitian generators $x^\mu$ as well as a
differential calculus $\Omega^*(\c{A})$ over $\c{A}$ which has certain
'parallelizability' properties; notably the $\c{A}$-module
$\Omega^1(\c{A})$ is free. This will be discussed in more detail
below. The $x^\mu$ will be referred to as `position generators'.  We
shall suppose also that there is a set of $n(=4)$ antihermitian
`momentum generators' $\lambda_\alpha$ and a `Fourier transform'
$$
F: x^\mu \longrightarrow \lambda_\alpha = F_\alpha (x^\mu)
$$
which takes the position generators to the momentum generators.
Let $\rho$ be a representation of $\c{A}$ as an algebra of linear
operators on some Hilbert space.  For every $k_\mu \in \b{R}^4$ one
can construct a unitary element $u(k) = e^{ik_\mu x^\mu}$ of $\c{A}$ 
and one can consider the weakly closed algebra $\c{A}_\rho$ generated
by the image of the $u(k)$ under $\rho$.  The momentum operators
$\lambda_\alpha$ are also unbounded but using them one can construct
also a set of `translation' operators 
$\hat u(\xi) = e^{\xi^\alpha \lambda_\alpha}$ whose image under $\rho$
belongs also to $\c{A}_\rho$. In general $\hat u u \neq u \hat u$; if
the metric which we introduce is the flat metric then we shall see
that $[\lambda_\alpha, x^\mu] = \delta^\mu_\alpha$ and in this case we
can write the commutation relations as $\hat u u = q u \hat u$ with 
$q = e^{i k_\mu \xi^\mu}$;  the `Fourier transform' is the simple linear
transformation
\begin{equation}
\lambda_\alpha  = \frac{1}{i\kbar}\theta^{-1}_{\alpha\mu} x^\mu   \label{bc}
\end{equation}
for some symplectic structure $\theta^{\alpha\mu}$. As a measure of
noncommmutativity, and to recall the many parallelisms with quantum
mechanics, we use the symbol $\kbar$, which will designate the square
of a real number whose value could lie somewhere between the Planck
length and the proton radius. 

Using $\lambda_\alpha$ we construct the derivation
$$
e_\alpha f = [\lambda_\alpha f].
$$
That is, the derivations are related to the momenta as
usual.  In the flat case mentioned above we find that
$$
e_\alpha x^\nu = 
\frac{1}{i\kbar}\theta^{-1}_{\alpha\mu} [x^\mu, x^\nu] = \delta_\alpha^\nu
$$
provided we introduce the commutation relations
$$
[x^\mu, x^\nu] = i\kbar \theta^{\mu\nu}
$$
or equivalently the Poisson structure 
$$
\{x^\mu, x^\nu\} = 
\lim_{\kbar\to 0}\frac 1{i\kbar} [x^\mu, x^\nu] = \theta^{\mu\nu}
$$
in the commutative limit.  

Since the center is trivial the algebra can also be considered as the
`phase space' and $F$ will be (implicitly) assumed to be an
automorphism of $\c{A}$:
$$
\widehat{u_1 \wm u_2} = \hat u_1 \hwm \hat u_2.
$$
We designate here for a moment the product with a star so we can
place a hat on it. In the commutative limit $F$ is an extension to phase
space of the ordinary Fourier transform.  As a simple example, one can
consider the two Pauli matrices $(x^\mu) = (\sigma^1,\sigma^2)$ as
`coordinates' for the algebra of $2\times 2$ matrices and the linear
combination 
$(\lambda_\alpha) = (-\sigma_2,\sigma_1) = \epsilon_{\mu\nu} \sigma^\nu$ 
as `momenta'. The derivations $e_\alpha = \ad \lambda_\alpha$ almost
behave as partial derivatives: $e_\alpha x^\mu = \delta^\mu_\alpha \sigma^3$. 
In this case there is no commutative limit.

In general the representation is not necessarily irreducible but we
shall assume that it is a direct sum of an arbitrary number of copies
of some representation which is irreducible.  The subalgebra
$\c{A}^\prime_\rho$ of those elements algebra $\c{B}(\c{H})$ of all
bounded operators on $\c{H}$ which commute with $\c{A}_\rho$ is
therefore in general nontrivial.  There does not seem to be a direct
role to be played by the $\c{A}^\prime_\rho$ except for the constraint
that it must grow as the algebra becomes commutative if this limit is
not to be singular.  Typically then, as the algebra becomes
increasingly commutative the representation becomes increasingly
reducible.

We shall suppose that $\c{A}$ has a commutative limit which is an
algebra $\c{C}(V)$ of smooth functions on a space-time $V$ endowed
with a globally defined moving frame $\theta^\alpha$ and thus a
metric.  By parallelizable we mean that the module $\Omega^1(\c{A})$
has a basis $\theta^\alpha$ which commutes with the elements of
$\c{A}$. For all $f \in \c{A}$
\begin{equation}
f\theta^\alpha = \theta^\alpha f.                          \label{f-theta}
\end{equation}
This is the correspondence principle we use. We shall see that it
implies that the metric components must be constants, a condition
usually imposed on a moving frame. The frame $\theta^\alpha$ allows
one~\cite{Mad00c} to construct a representation of the differential
algebra from that of $\c{A}$. The frame elements belong in fact to the
algebra $\c{A}^\prime_\rho$. Following strictly what one does in
ordinary geometry, we shall introduce the set of derivations
$e_\alpha$ to be dual to the frame $\theta^\alpha$, that is with
$$
\theta^\alpha(e_\beta) = \delta^\alpha_\beta.
$$ 

We define the differential exactly as did E.~Cartan in the commutative
case. If $e_\alpha$ is a derivation of $\c{A}$ then for every element 
$f \in \c{A}$ we define $df$ by the constraint 
$df(e_\alpha) = e_\alpha f$. The construction of the differential
calculus from the structure of the resulting module of 1-forms is
essentially due to Connes~\cite{Con94}; in the present notation it is
to be found elsewhere~\cite{Mad00c}.  It follows from general
arguments that the momenta $\lambda_\alpha$ must satisfy the
consistency condition~\cn{DimMad96,MadMou98}
\begin{equation}
2 \lambda_\gamma \lambda_\delta P^{\gamma\delta}{}_{\alpha\beta} - 
\lambda_\gamma F^\gamma{}_{\alpha\beta} - 
K_{\alpha\beta} = 0.                                           \label{consis}
\end{equation}
The $P^{\gamma\delta}{}_{\alpha\beta}$ define the product $\pi$ in the
algebra of forms:
\begin{equation}
\theta^\alpha \theta^\beta = P^{\alpha\beta}{}_{\gamma\delta} 
\theta^\gamma \theta^\delta.                                  \label{struc}
\end{equation}
This product is defined to be the one with the least relations which
is consistent with the module structure of the 1-forms.  The
$F^\gamma{}_{\alpha\beta}$ are related to the 2-form $d\theta^\alpha$
through the structure equations:
$$
d\theta^\alpha = - {1\over 2}C^\alpha{}_{\beta\gamma} 
\theta^\beta \theta^\gamma.
$$
In the noncommutative case the structure elements are defined as
\begin{equation}
C^\alpha{}_{\beta\gamma} = F^\alpha{}_{\beta\gamma} - 
2 \lambda_\delta P^{(\alpha\delta)}{}_{\beta\gamma}.              \label{s-e}
\end{equation}
It follows that
\begin{equation}
e_\alpha C^\alpha{}_{\beta\gamma} = 0.                           \label{fix}
\end{equation}
This must be imposed then at the commutative level and can be used as a 
gauge-fixing condition. 

Finally, to complete the definition of the coefficients of the
consistency condition~(\ref{consis}) we introduce the special 1-form 
$\theta = -\lambda_\alpha\theta^\alpha$. In the commutative, flat limit
$$
\theta \rightarrow i\partial_\alpha dx^\alpha.
$$
It is referred to as a `Dirac operator'~\cite{Con94}. As an
(antihermitian) 1-form $\theta$ defines a covariant derivative on an
associated $\c{A}$-module with local gauge transformations given by
the unitary elements of $\c{A}$.  The $K_{\alpha\beta}$ are related to
the curvature of $\theta$:
$$
d\theta + \theta^2 = - {1\over 2} K_{\alpha\beta} \theta^\alpha \theta^\beta.
$$
All the coefficients lie in the center $\c{Z}(\c{A}_\kbar)$ of the algebra.
With no restriction of generality we can impose the conditions
\begin{equation}
C^\epsilon{}_{\gamma\delta} = 
P^{\alpha\beta}{}_{\gamma\delta} C^\epsilon{}_{\alpha\beta}, \qquad 
K_{\gamma\delta} = 
P^{\alpha\beta}{}_{\gamma\delta} K_{\alpha\beta}.              \label{C-con}
\end{equation}

Equation~(\ref{s-e}) is the correspondence principle which associates
a differential calculus to a metric.  On the left in fact the quantity
$C^\alpha{}_{\beta\gamma}$ determines a moving frame, which in turn
fixes a metric; on the right are the elements of the algebra which fix
to a large extent the differential calculus. A `blurring' of a
geometry proceeds via this correspondence. It is evident that in the
presence of curvature the 1-forms cease to anticommute. On the other
hand it is possible for flat `space' to be described by `coordinates'
which do not commute.  

The correspondence principle between the commutative and
noncommutative geometries can be also described as the map
\begin{equation}
\t{\theta^\alpha}  \mapsto \theta^\alpha                      \label{cp}
\end{equation}
with the product satisfying the condition
\begin{equation}
\t{\theta^\alpha} \t{\theta^\beta} \mapsto 
P^{\alpha\beta}{}_{\gamma\delta} \theta^\gamma \theta^\delta      \label{cp0}
\end{equation}
The tilde on the left is to indicate that it is the commutative form.
The condition can be written also as
$$
\t{C}^\alpha{}_{\beta\gamma} \mapsto 
C^\alpha{}_{\eta\zeta} P^{\eta\zeta}{}_{\beta\gamma}
$$
or as 
$$
\lim_{\kbar\to 0}C^\alpha{}_{\beta\gamma} = \t{C}^\alpha{}_{\beta\gamma}.
$$
A solution to these equations would be a solution to the problem we
have set. It would be however unsatisfactory in that no smoothness
condition has been imposed. This can at best be done using the inner
derivations. We shall construct therefore the set of momentum
generators. The procedure we shall follow is not always valid; a
counter example has been constructed~\cite{MadSae98} for the flat
metric on the torus.

We write $P^{\alpha\beta}{}_{\gamma\delta}$ in the form
\begin{equation}
P^{\alpha\beta}{}^{\phantom{\alpha\beta}}_{\gamma\delta} = 
\frac 12 \delta^{[\alpha}_\gamma \delta^{\beta]}_\delta +
i\kbar Q^{\alpha\beta}{}^{\phantom{\alpha\beta}}_{\gamma\delta}.     \label{P}
\end{equation}
It will be convenient also to introduce the notation
\begin{equation}
I_{\alpha\beta} = 
K_{\alpha\beta} + \lambda_\gamma I^\gamma{}_{\alpha\beta}, \qquad
I^\gamma{}_{\alpha\beta} = F^\gamma{}_{\alpha\beta} - 
2i\kbar\lambda_\delta Q^{\gamma\delta}{}_{\alpha\beta}          \label{Gab}
\end{equation}
so we can write~(\ref{consis}) in the form
$$
[\lambda_\alpha, \lambda_\beta] = I_{\alpha\beta} -
i\kbar[\lambda_\gamma, \lambda_\delta]
Q^{\gamma\delta}{}_{\alpha\beta}.                           
$$
If we replace the commutator on the right-hand side by its value we obtain
the equation
$$
[\lambda_\alpha, \lambda_\beta] = I_{\alpha\beta} -
i\kbar Q^{\gamma\delta}{}_{\alpha\beta}
(K_{\gamma\delta} + I_{\gamma\delta}) + (i\kbar)^2
Q^{\eta\zeta}{}_{\alpha\beta}  Q^{\gamma\delta}{}_{\eta\zeta}
[\lambda_\gamma, \lambda_\delta].
$$
This can be simplified by a redefinition 
$$
K_{\alpha\beta} - i\kbar Q^{\gamma\delta}{}_{\alpha\beta} K_{\gamma\delta}
\mapsto K_{\alpha\beta}, \qquad
F^\epsilon{}_{\alpha\beta} - 
i\kbar Q^{\gamma\delta}{}_{\alpha\beta} F^\epsilon{}_{\gamma\delta}
\mapsto F^\epsilon{}_{\alpha\beta}
$$
of the coefficients $K_{\alpha\beta}$ and $F^\gamma{}_{\alpha\beta}$.
Iterating this proceedure one obtains the consistency condition in the 
form
\begin{equation}
[\lambda_\alpha, \lambda_\beta] = I_{\alpha\beta}            \label{consis2}
\end{equation}
as well as the condition
$$
Q^{[\gamma\delta]}{}_{\alpha\beta} = 0 
$$
on the coefficients. The condition that
$P^{\gamma\delta}{}_{\alpha\beta}$ be an idempotent now implies a
similar condition on $i\kbar Q^{(\gamma\delta)}{}_{(\alpha\beta)}$.
Unless we are in a non-perturbative regime we can set then the latter
equal to zero.

We introduce the derivations
$e_{\alpha\beta} = \ad (\lambda_\gamma \lambda_\delta)$
so that one can write the identity
\begin{equation}
i\kbar e_{\alpha\beta} = i\kbar \lambda_{(\alpha} e_{\beta)} - 
i\kbar e_\beta e_\alpha.                                        \label{sc}
\end{equation}
The first term on the right remains in the commutative limit as a
vector field; the second term vanishes. The second term resembles a
second derivative, but as a derivation. It is easy to see here the
relation between an expression from noncommutative geometry and an
effective commutative limit with higher-order derivative corrections. 

In the commutative case a moving frame is dual to a set of derivations
$\t{e}_\alpha$ which satisfy the commutation relations
\begin{equation}
[\t{e}_\alpha, \t{e}_\beta] = 
\t{C}^\gamma{}_{\alpha\beta}\t{e}_\gamma             \label{te-te}
\end{equation}
If we use the correspondence principle and the
expression~(\ref{s-e}) we find that this becomes
\begin{equation}
[e_\alpha, e_\beta] = F^\gamma{}_{\alpha\beta} e_\gamma -
2i\kbar Q^{\gamma\delta}{}_{\alpha\beta} e_{\gamma\delta}  
= F^\gamma{}_{\alpha\beta} e_\gamma -
2i\kbar Q^{\gamma\delta}{}_{\alpha\beta} 
(\lambda_{(\gamma} e_{\delta)} - e_{\gamma\delta})              \label{sc2}
\end{equation}
The last term on the right-hand side vanishes in the commutative limit.

The Jacobi identity with two momenta and one position generator can be
written as
\begin{equation}
e_{[\alpha} e^\mu_{\beta]} = 
e^\mu_\gamma I^\gamma{}_{\alpha\beta}.                           \label{G-C}
\end{equation}
Because of the symmetries of the indices $I_{\alpha\beta}$ can be
considered as a field strength for a gauge potential which takes its
values in the formal Lie algebra associated to the group of unitary
elements of the associative algebra $\c{A}$. Although the algebra is
generally of infinite dimension the relation~(\ref{consis}) acts as a
finiteness condition and one can consider the $\lambda_\alpha$ as
generators of a Lie algebra with commutation relations
given~\cite{DimMad96,JurMolSchSchWes01} by the
bracket~(\ref{consis2}). It is convenient to introduce the (right) dual
$$
Q^{*\zeta\eta\gamma\delta} =
\frac{1}{2} \epsilon^{\alpha\beta\gamma\delta}
Q^{\zeta\eta}{}_{\alpha\beta}
$$
of $Q^{\zeta\eta}{}_{\alpha\beta}$ and to define the Jacobi anomaly
for three momenta as
$$
A_J^\delta = \frac 14\epsilon^{\alpha\beta\gamma\delta}
[\lambda_\alpha,[\lambda_\beta,\lambda_\gamma]].
$$
The Jacobi identities can be written then as
\begin{equation}
A_J^\alpha = 0.                                              \label{J3}
\end{equation}
With the Ansatz we are using we find that
$$
A_J^\alpha =
-2i\kbar K_{\beta\gamma}Q^{*\beta\delta\gamma\alpha}\lambda_\delta
+ (-2i\kbar)^2 Q^{(\delta\gamma}{}_{\beta\gamma}
Q^{*\beta\eta)\gamma\alpha} \lambda_\delta\lambda_\gamma\lambda_\eta.
$$
Equation~(\ref{J3}) must be verified in each example. 
Because of the relations amongst the momenta, sufficient but not
necessary conditions for the left-hand side to vanish are
\begin{eqnarray}
K_{\beta\gamma}Q^{*\beta\delta\gamma\alpha} = 0,          \label{jac1}\\[6pt]
Q^{(\delta\gamma}{}_{\beta\gamma}
Q^{*\beta\eta)\gamma\alpha} = 0.                        \label{jac2}
\end{eqnarray}
In the example we shall find that both these equations place
restrictions on the coefficients.

It is necessary~\cite{DubMadMasMou96b} to introduce a flip operation
$$
\sigma:\;\Omega^1(\c{A}_\kbar) \otimes \Omega^1(\c{A}_\kbar) \rightarrow
\Omega^1(\c{A}_\kbar) \otimes \Omega^1(\c{A}_\kbar)
$$
to define the reality condition and the Leibniz rules.  If we write
$$
S^{\alpha\beta}{}_{\gamma\delta} = \delta^\beta_\gamma \delta^\alpha_\delta
+ i\kbar T^{\alpha\beta}{}_{\gamma\delta}
$$
we find that a choice~\cn{DimMad96} of connection which is
torsion-free, and satisfies all Leibniz rules is given by
\begin{equation}
\omega^\alpha{}_{\beta} = \frac 12 F^\alpha{}_{\gamma\beta} \theta^\gamma
+ i\kbar \lambda_\gamma T^{\alpha\gamma}{}_{\delta\beta}\theta^\delta.
                                                               \label{t-free}
\end{equation}
We can compare this with the expression~(\ref{s-e}) for the structure
elements.  

The relation
$$
\pi \circ (1 + \sigma) = 0
$$
must hold~\cn{DimMad96,MadMou98} to assure that the torsion be a
bilinear map. In terms of the coefficients
$P^{\alpha\beta}{}_{\gamma\delta}$ it can be written in the form
$$
T^{\alpha\beta}{}_{\eta\zeta} P^{\eta\zeta}{}_{\gamma\delta} +  
Q^{(\alpha\beta)}{}_{\gamma\delta} = 0.  
$$
The symmetric part $T^{(\alpha\beta)}{}_{(\gamma\delta)}$, which is here 
arbitrary, will be fixed by the condition that the connection be metric.
It follows that
\begin{equation}
\omega^\alpha{}_{\eta\zeta}P^{\eta\zeta}{}_{\beta\gamma} =
\frac 12 C^\alpha{}_{\beta\gamma}.                     \label{T0}
\end{equation}
This is the usual relation between the Ricci-rotation coefficients
and the Levi-Civita connection. Using it one can deduce~(\ref{s-e})
from~(\ref{t-free}).  Using the vanishing-torsion condition one can
write the expression for the commutator of the derivations as
\begin{equation}
[e_\alpha, e_\beta] - \omega^\gamma{}_{[\alpha\beta]} e_\gamma = 
- 2i\kbar Q^{\gamma\delta}{}_{\alpha\beta} 
(e_{\gamma} e_{\delta} - \omega^\epsilon{}_{\gamma\delta} e_\epsilon).
                                                         \label{consis3}
\end{equation}
A second form of the correspondence principle is the map
\begin{equation}
\t{\omega}^\alpha{}_{\beta} \t{\theta} \mapsto 
\omega^\alpha{}_{\beta} \theta                             \label{cp2}
\end{equation}
between the commutative and noncommutative connection forms.

We shall suppose that $\c{A}_\kbar$ has a metric, which we define as a
bilinear map 
\begin{equation}
g:\; \Omega^1(\c{A}_\kbar) \otimes \Omega^1(\c{A}_\kbar) 
\rightarrow \c{A}_\kbar.                                        \label{metric} 
\end{equation} 
In terms of the frame one can define the metric by the condition that
\begin{equation}
g(\theta^\alpha \otimes \theta^\beta)= g^{\alpha\beta}.          \label{f-m}
\end{equation}
The $g^{\alpha\beta}$ are taken to be deformed versions 
$g^{\alpha\beta} \mapsto g^{\alpha\beta} + i \kbar h^{\alpha\beta}$
of the components of the standard euclidean or lorentzian metric
$g^{\alpha\beta}$ on $\b{R}^n$ which satisfy~\cn{FioMad98a} the
symmetry condition
\begin{equation}
P^{\alpha\beta}{}_{\gamma\delta} (g+i\kbar h)^{\gamma\delta} = 0.   
                                                             \label{symmet}
\end{equation}
Together with the reality condition
$$
(g+i\kbar h)^{\beta\alpha} + i\kbar T^{\alpha\beta}{}_{\gamma\delta}
(g+i\kbar h)^{\gamma\delta} = ((g+i\kbar h)^{\beta\alpha})^*
$$
this means that in fact $h^{\alpha\beta} = 0$ and also that the flip
satisfies the constraint
$$
T^{\alpha\beta}{}_{\gamma\delta}g^{\gamma\delta} = 0.
$$
We suppose then that the coefficients of the metric are the usual
commutative ones. The various reality
conditions~\cite{Con95,FioMad98a} imply also that
$$
(Q^{\alpha\beta}{}_{\gamma\delta})^* =
Q^{\alpha\beta}{}_{\gamma\delta} + o(i\kbar), \qquad
(T^{\alpha\beta}{}_{\gamma\delta})^* =
T^{\alpha\beta}{}_{\gamma\delta} + o(i\kbar).
$$

The connection is compatible with the metric if
\begin{equation}
T^{\alpha\gamma}{}_{\delta\epsilon} g^{\epsilon\beta} +
T^{\beta\gamma}{}_{\delta\epsilon} g^{\alpha\epsilon} +
i\kbar T^{\beta\gamma}{}_{\epsilon\zeta} g^{\eta\zeta}
T^{\alpha\epsilon}{}_{\delta\eta} = 0.                          \label{m-c3}
\end{equation}
To first order this simplifies to the usual condition
\begin{equation}
T^{(\alpha\gamma}{}_{\delta}{}^{\beta)} = o(i\kbar).            \label{m-c2}
\end{equation}
The index was lifted here with the metric.  

If we introduce formally the `covariant derivative' $D_\alpha X^\beta$
of a `vector' $X^\alpha$ by the formula
$$
D_\alpha X^\beta = \omega^\beta{}_{\alpha\gamma} X^\gamma
$$
and the covariant derivative $D_\alpha (X^\beta Y^\gamma)$ of the
product of two such `vectors' as
$$
D_\alpha (X^\beta Y^\gamma) = D_\alpha X^\beta Y^\gamma + 
S^{\beta\delta}{}_{\alpha\epsilon} X^\epsilon D_\delta Y^\gamma,
$$
since there is a `flip' as the index on the derivation crosses the
index on the first `vector', then the condition~(\ref{m-c3}) that the
connection be metric can be written
$$
D_\alpha g^{\beta\gamma} = 0
$$
as usual. More satisfactory origins of the condition~(\ref{m-c3})
can be given~\cite{DubMadMasMou95}.

The `position' commutation relations are of the form
\begin{equation}
[x^\mu, x^\nu] = i\kbar J^{\mu\nu}.                        \label{x-x}
\end{equation}
The right-hand side is not necessarily in the center.  The system of
operator equations we must solve consists of Equation~(\ref{s-e}),
Equation~(\ref{consis2}), the metric compatibility
equation~(\ref{m-c2}) as well as the constraints on the coefficients
given above.  Using the same formal manipulations as above we can find
the commutation relations for the position generators as solution to a
set of differential conditions. From~(\ref{x-x}) we see that
\begin{equation}
i\kbar e_\gamma J^{\mu\nu} = [x^{[\mu}, e^{\nu]}_\gamma]  \label{pxx}.
\end{equation}
We choose the symplectic structure $\theta^{\alpha\beta}$ such that
$$
\theta^{-1}_{\alpha\beta} = -i\kbar K_{\alpha\beta} + 
o(\omega^\gamma_{\alpha\beta})
$$ 
and such that 
\begin{equation}
J^{\mu\nu} = 
e^\mu_\alpha e^\nu_\beta \theta^{\alpha\beta}.            \label{J-theta}
\end{equation}
This cannot be generally possible unless
$\theta^{(\alpha\beta)}\neq 0$.  It will allow us to define
(formally) a divergence $D_\mu J^{\mu\nu}$ by setting
\begin{equation}
D_\mu J^{\mu\nu} = 
e^\nu_\beta D_\alpha \theta^{\alpha\beta}           \label{div}
\end{equation}
and using the connection form~(\ref{t-free}).
The condition 
$$
D_\mu J^{\mu\nu} =0 
$$
establishes a link~\cite{BouCah99} between the symplectic structure
and the connection. We cannot make any general statements concerning
this condition in the present context.

It is tempting to suppose that to lowest order at least, in a
semiclassical approximation, there is an analogue of Darboux's lemma
and that it is always possible to choose generators which satisfy
commutation relations of the form~(\ref{x-x}) with the right-hand in
the center.  However the example we shall examine in detail shows that
this is not always the case. Having fixed the generators, the
manifestations of curvature would be found then in the form of the
frame. The two sets of generators $x^\mu$ and $\lambda_\alpha$
satisfy, under the assumptions we make, three sets of equations.  The
commutation relations~(\ref{x-x}) for the position generators $x^\mu$
and the associated Jacobi identities permit one definition of the
algebra. The commutation relations for the momentum generators permit
a second definition. The conjugacy relations assure that the two
descriptions concern the same algebra.  We shall analyze these
identities later using the example to show that they have interesting
non-trivial solutions.

There are two sets of equations to be solved simultaneously in the
four variables $\lambda_\alpha$. The first set is the
set~(\ref{s-e}), a system of $(4-1)\times 6 = 18$ equations which are
to be read now from right to left as equations in the $\lambda_\alpha$
with given `sources' $C^\alpha{}_{\beta\gamma}$. This gives the
Fourier transform as a functional of the gravitational field. The
boundary conditions are obscure but could be taken, for lack of
something better, as Equations~(\ref{bc}); this is then to be the form of
the map from $x^\mu$ to $\lambda_\alpha$ when there is no source.
The second set of equations is the set of consistency
conditions~(\ref{consis2}). In the commutative limit the Fourier
transformation becomes a set of four functionals, which could be considered
as a special coordinate transformation, and in this limit the
consistency conditions give rise to differential equations. In general
however the transformation will not be a local one, a simple point to
point map, but a more elaborate morphism of the entire algebra of
functions into itself. It would be analogous to the unitary operator
of field theory which takes the values of the fields at minus infinity
to their values at some fixed finite time. The former would be the
$\lambda_\alpha$ and the latter the $x^\mu$. Since the algebra is
noncommutative one might suppose this morphism to be also inner and
write its inverse as
$$
x^\mu \rightarrow \lambda_\alpha =
\frac{1}{i\kbar} \theta^{-1}_{\alpha\mu} U^{-1} x^\mu U 
$$
for some $U$. The map defined by Equations~(\ref{bc}) is not however
of this form. We shall return to this briefly in Section~\ref{Heis}.

All this can be clarified somewhat by repeating it in the linear
approximation, by which we mean a Fourier transformation `close' to
that given by~(\ref{bc}). Equation~(\ref{s-e}) can be written as
$$
C^\alpha{}_{\beta\gamma}(x^\mu) = F^\alpha{}_{\beta\gamma} -
4i\kbar\lambda_\delta Q^{\alpha\delta}{}_{\beta\gamma}.
$$
To the lowest order the right-hand side is linear in the metric
components and the Fourier map is linear. This is a simple problem in
linear algebra and easy to solve, but the solution might be the
trivial one. Beyond the linear approximation the complication is due
mainly to the fact that, although the right-hand side remains linear
in $\lambda_\alpha$, the left-hand side becomes very complicated
because of the Fourier transformation. The $C^\alpha{}_{\beta\gamma}$
are given functions of $x^\mu$ and in general simple; as functions of
the $\lambda_\alpha$ they are quite complicated.

A commutative gauge transformation is an element of the algebra
$M_4(\c{C}(\b{R}^4))$.  The product in the algebra of forms is not
necessarily anti-commutative but the extra terms will not appear in
the expression for the differential of a form, because of the
relation~(\ref{C-con}) and the form of the Ansatz we are using. We
shall adopt the rule that the gauge transformation must be made before
`quantization', not after. This is for many reasons, principally the
fact that the `after-quantization' gauge
`group'~\cite{JurMolSchSchWes01} is more difficult to define.

We start then with a given frame and associated rotation coefficients
and we transform the frame to a new one, to whose rotation
coefficients we give the name $C^{\prime\alpha}{}_{\beta\gamma}$.  As an
additional simplification we shall calculate all quantities retaining
only terms first order in the gauge elements.  The rotation
coefficients are then given by
$$
C^{\prime\alpha}{}_{\beta\gamma} = C^\alpha{}_{\beta\gamma} + 
D_{[\beta} H^\alpha_{\gamma]}
$$
for some first-order gauge transformation whose form must be found
as part of the problem. We have here introduced the notation
$$
D_{\beta} H^\alpha_{\gamma} =
e_{\beta} H^\alpha_{\gamma} + [C,H]^\alpha{}_{\beta\gamma}, \qquad
[C,H]^\alpha{}_{\beta\gamma} = 
C^\alpha{}_{\beta\delta} H^\delta_{\gamma} -
H^\alpha_{\delta} C^\delta{}_{\beta\gamma} 
$$
The use we make of the letter $H$ is motivated in
Section~\ref{Heis}. We have chosen the original frame more or less
arbitrarily. It will determine a symplectic form and a metric and in
general it must be gauged so that these two objects are correctly
related. The metric is of course Lorentz invariant but the symplectic
form not.

From~(\ref{s-e}) one finds that
\begin{equation}
C^\alpha{}_{\beta \gamma} + D_{[\beta} H^\alpha_{\gamma]}
= F^\alpha{}_{\beta \gamma} -
4i\kbar \lambda_\delta Q^{\alpha\delta}{}_{\beta \gamma}.        \label{s-e2}
\end{equation}
Only the position generators $x^\mu$ appear on the left-hand side
and only the momentum generators $\lambda_\pm$ appear on the
right-hand side. Finally the structure equations can be written as
the commutation relations
\begin{equation}
[\lambda_\beta, \lambda_\gamma] = K_{\beta\gamma}
+ \frac 12 \lambda_\alpha (F^\alpha{}_{\beta \gamma} +
C^\alpha{}_{\beta \gamma} + D_{[\beta} H^\alpha_{\gamma]}).   \label{consis4}
\end{equation}
In the example we give we shall render more explicit this equation.
We could have introduced $\lambda_0 = 1$ and written the constant term
as $\lambda_0 F^0{}_{\alpha\beta} = K_{\alpha\beta}$. It could be
thought of as a spontaneous shift \`a la Brout-Englert-Higgs-Kibble in
the eigenvalues of the operators $\lambda_\alpha$, We must also deal
with the fact that we have imposed no associativity condition on the
product. For example the quantities $P^{\alpha\beta}{}_{\gamma\delta}$
or $S^{\alpha\beta}{}_{\gamma\delta}$ have not been required to
satisfy the Yang-Baxter equation. We refer to previous
work~\cite{FioMad98a} for a discussion of the problems this might
entrain concerning reality conditions within the present formalism.
The right-hand side of Equation~(\ref{consis4}) must at least satisfy
the Jacobi identities.  We shall focus attention on this point in the
example.

The problem of gauge invariance and the algebra of observables is a
touchy one upon which we shall not dwell. We notice in
Section~\ref{Heis} that formally a gauge transformation in
noncommutative geometry is the same as the time evolution in quantum
field theory. This is somewhat the analog of the result from mechanics
to the effect that the hamiltonian generates that particular
symplectic transformation which is the time evolution.

It is obvious that not all of the elements of $\c{A}$ are gauge
invariant but not that all observables are gauge-invariant.  A gauge
transformation is a local rotation
\begin{equation}
\theta^\alpha \mapsto \theta^{\prime\alpha} =
\Lambda^{-1}{}^\alpha_\beta \theta^\beta.            \label{lr}
\end{equation}
This formula is slightly misleading since we have written
$\theta^{\prime\alpha}$ as though it belonged to the module of 1-forms
of the original calculus whereas in fact it is a frame for a different
differential calculus. One should more properly use a direct-sum
notation and consider $\Lambda^\alpha_\beta$ and its inverse as maps
between the two terms; this would complicate considerably the formulae.
The structure elements transform classically as
$$
\t{C}^{\prime\alpha}{}_{\beta\gamma} = \t{\Lambda}^{-1}{}^\alpha_\delta
\t{C}^{\delta}{}_{\epsilon\zeta} \t{\Lambda}^\epsilon_\beta
\t{\Lambda}^\zeta_\gamma + 
\t{\Lambda}^{-1}{}^\alpha_\delta \t{e}_\epsilon
\t{\Lambda}^\delta_{[\gamma} \t{\Lambda}^\epsilon_{\beta]}.
$$
This is a familiar formula but it cannot be taken over as is to the
noncommutative case; under the gauge transformation the calculus
transforms also. This is apparent in the conditions~(\ref{C-con}).
Within the context of the present calculi the nearest one can come is
the transformation
$$
C^{\prime\alpha}{}_{\beta\gamma} = \Lambda^{-1}{}^\alpha_\delta
C^{\delta}{}_{\epsilon\zeta} \Lambda^\epsilon_\beta
\Lambda^\zeta_\gamma + \delta C^\alpha{}_{\beta\gamma} \qquad \delta
C^\alpha{}_{\beta\gamma} = - 2e_\delta\Lambda^{-1}{}^\alpha_\epsilon
\Lambda^\delta_\zeta \Lambda^\epsilon_\eta
P^{\prime\zeta\eta}{}_{\beta\gamma}.
$$
If $P$ and $P^\prime$ are of the form considered here one can write
\begin{eqnarray*}
&&C^{\prime\alpha}{}_{\beta\gamma} = \Lambda^{-1}{}^\alpha_\delta
C^{\delta}{}_{\epsilon\zeta} \Lambda^\epsilon_\beta
\Lambda^\zeta_\gamma - 2e_\delta\Lambda^{-1}{}^\alpha_\epsilon
[\Lambda^\delta_\zeta, \Lambda^\epsilon_\eta]
P^{\prime\zeta\eta}{}_{\beta\gamma}\\[6pt]
&&\phantom{C^{\prime\alpha}{}_{\beta\gamma} } +
\Lambda^{-1}{}^\alpha_\delta e_\epsilon \Lambda^\delta_{[\beta}
\Lambda^\epsilon_{\gamma]} + \Lambda^{-1}{}^\alpha_\delta e_\epsilon
\Lambda^\delta_\zeta \Lambda^\epsilon_\eta
P^{\prime(\zeta\eta)}{}_{\beta\gamma}.
\end{eqnarray*}
In the linear approximation, with $\Lambda^\alpha_\beta \simeq
\delta^\alpha_\beta + H^\alpha_\beta$, the inhomogeneous
term simplifies to the expression
\begin{equation}
- 2e_\delta\Lambda^{-1}{}^\alpha_\epsilon \Lambda^\delta_\zeta
\Lambda^\epsilon_\eta P^{\prime\zeta\eta}{}_{\beta\gamma} = e_{[\beta}
H^\alpha_{\gamma]} + e_\delta
H^\alpha_\epsilon P^{(\delta\epsilon)}{}_{\beta\gamma}.
                                                                \label{SW1}
\end{equation}        
Since
$$
\theta^{\prime\alpha}(e_\beta) = \Lambda^{-1}{}^\alpha_\beta
$$
in a certain formal way one can imagine that
$$
e^\prime_\alpha \sim \Lambda^\beta_\alpha e_\beta
$$
except that the right-hand-side is not a derivation in general. It
makes sense however to introduce 
$\bar{\lambda}^\prime_\alpha = \Lambda_\alpha^\beta \lambda_\beta$ 
and to compare it with the elements $\lambda^\prime_\beta$ defined to
be those which generate the derivations dual to
$\theta^{\prime\alpha}$.  This means that we must consider the
identity
$$
\Lambda^{-1}{}^\gamma_\beta \theta^\beta_\mu
[\lambda^\prime_\alpha, x^\mu] = \theta^\gamma_\mu
[\Lambda^{-1}{}^\beta_\alpha \bar{\lambda^\prime}_\beta, x^\mu]
$$
as an equation for $e^\prime_\alpha$ in terms of 
$\bar {e}^\prime_\alpha = \ad \bar{\lambda}^\prime_\alpha$ and solve. There
results a rather complicated situation because of the change in
Poisson structure which must accompany a gauge transformation in
general. We shall attempt to elucidate this with the example.

\initiate
\section{The commutative plane wave}\label{spp}

Plane-fronted gravitational waves with parallel rays ($pp$-waves) have
a metric~\cite{KraSteMacHer80} which can be
defined~\cite{DesPirRob76} by the moving frame $\t{\theta}^a = dx^a$,
$x^a = (x, y)$ and
\begin{equation}
\t{\theta}^+ = \frac 1{\sqrt 2} (\t{\theta}^0 + \t{\theta}^3) = 
\t{\Lambda}^{-1} (du + \t{h} dv ), \qquad 
\t{\theta}^- = \frac 1{\sqrt 2} (\t{\theta}^0 - \t{\theta}^3)
= \t{\Lambda} dv.                                        \label{th-cl}
\end{equation}
The generators of $\c{A}$ will be chosen $x^\mu = (u,v,x,y)$, with $u$
and $v$ null coordinates in the commutative limit.  The $\t{\Lambda}$
is for the moment an arbitrary invertible element of $\c{A}$.  When
$\t{h}=0$ the space-time is Minkowski and the vector $\partial_v$ is
tangent to the bicharacteristics along the characteristic surfaces $ u
= \mbox{constant}$.  If $\t{h} = \t{h}(v)$ one can introduce an
element $I\t{h}(v) \in \c{A}$ such that $\partial_v I\t{h}(v) =
\t{h}(v)$ and use it to define $u^\prime = u + I\t{h}(v)$; the
space-time is still Minkowski.  With $\t{\Lambda} = 1$ then one has
$\t{\theta}^+ = du^\prime$ and $\t{\theta}^- = dv$ The map defined by
$(uv) \mapsto (u^\prime,v)$ does not respect the structure of $\c{A}$
as an algebra but it does leave invariant the commutator: $[u^\prime,
v] = [u,v]$.  Consider the derivations $\t{e}_\alpha$ dual to the
moving frame. In the commutative limit we have $\t{e}_1 = \partial_x$,
$\t{e}_2 = \partial_y$ and
$$
\t{e}_+ = \t{\Lambda}\partial_u, \qquad 
\t{e}_- = \t{\Lambda}^{-1}(\partial_v - \t{h} \partial_u).
$$
The metric-compatible connection form is given by
$$
\t{\omega}^\alpha{}_\beta = 
\left(\begin{array}{ccc} \t{\omega}^+{}_+ &0
&\t{\omega}^+{}_b \\[4pt] 
0 &\t{\omega}^-{}_- &0 \\[4pt] 0
&\t{\omega}^a{}_- &0  
\end{array}\right)
$$
with
$$
\t{\omega}^+{}_+ = 
\t{\Lambda}^{-2}\t{e}_+ \t{h} \t{\theta}^- + 
\t{\Lambda}^{-1} d\t{\Lambda}, \qquad
\t{\omega}^+{}_a = \t{\Lambda}^{-1} \t{e}_a \t{h}  \t{\theta}^-.
$$
One of the expressions of the zero-torsion condition is the identity
$$
\t{D}_{[\alpha}\t{D}_{\beta]} \t{x}^\mu = 0
$$
which can also be written as
\begin{equation}
\t{e}_{[\alpha}\t{e}_{\beta]} \t{x}^\mu = 
\t{\omega}^\gamma{}_{[\alpha\beta]} \t{e}_\gamma x^\mu.       \label{com}
\end{equation}
This is the same as Equation~(\ref{te-te}.

When $\tilde{\Lambda}=1$ the moving frame satisfies the commutation
relations
$$
\begin{array}{ll}
[\t{e}_+, \t{e}_-] = - \t{e}_+ \t{h} \t{e}_+, &
[\t{e}_-, \t{e}_a] = \t{e}_a \t{h} \t{e}_+, \\[4pt]
[\t{e}_+, \t{e}_a] = 0, &
[\t{e}_a, \t{e}_b] = 0.
\end{array}
$$
If one includes a possible gauge transformation $\t{\Lambda}$ then one
finds that the non-zero structure coefficients are given by
\begin{equation}
\begin{array}{l}
\t{C}^+{}_{a -} = -\t{\Lambda}^{-2}\t{e}_a \t{h},\\[4pt]
\t{C}^+{}_{+ -} = -\t{\Lambda}^{-2}\t{e}_+ \t{h}
- \t{\Lambda}^{-1}\t{e}_-\t{\Lambda}          ,\\[4pt]
\t{C}^\pm{}_{\alpha \pm} = \pm\t{\Lambda}^{-1}\t{e}_\alpha\t{\Lambda}.
                                                                \label{Cclas} 
\end{array}
\end{equation}
In writing this we have used the fact that the connection form
commutes with the coordinates. This will no longer be true in general
and extra terms can appear.

\initiate
\section{$pp$-algebras}

The noncommutative generalization will involve essentially finding an
equivalent of Equation~(\ref{te-te}). It will be a commutation
relation of the form~(\ref{consis3}) and better expressed in terms of
the momentum generators. This will fix the algebra as well the
calculus. That is, the correspondence principle which we shall
actually use is a modified version of the map
$$
\t{e}_\alpha \mapsto \lambda_\alpha
$$
which is, of course, that introduced by Bohr.  This representation
is valid only in the limit $\kbar \to \infty$.  The commutation
relations for the $\t{e}_\alpha$ must be `lifted' to commutation
relations between the $\lambda_\alpha$. For this we notice that the
$L^2$-completion of the algebra of functions on the space-time can be
identified with $\c{H} = L^2(\b{R}^4)$. We define then for smooth
$\psi \in \c{H}$
\begin{equation}
\lambda_\alpha \psi = \t{e}_\alpha \psi.                 \label{repres}
\end{equation}
Once we have implemented the correspondence principle then this set of
momentum generators will satisfy the consistency
Equation~(\ref{consis}), however with $K_{\alpha\beta} = 0$. To obtain a
non-zero value for the central extension one must modify the
representation. We do this in our analysis of another
metric~\cite{MacMadZou02}; here we restrict our considerations to formal
algebras. 

According to the correspondence principle as expressed by~(\ref{cp})
we introduce a frame $\theta^\alpha$ which has the commutative frame as
a limit and is as `near' to it in form as possible for all values of
$\kbar$. We shall assume in this example that the noncommutative frame
has the same functional form as its limit.  When $h=0$, we have noticed,
the space-time is flat; the noncommutative version is also flat. 
The frame is the noncommutative form of~(\ref{th-cl}) which we write in
the same way as
\begin{equation}
\theta^+ = \frac 1{\sqrt 2}(\theta^0 + \theta^3) = 
\Lambda^{-1}(du + h dv ), \qquad 
\theta^- = \frac 1{\sqrt 2} (\theta^0 - \theta^3) = 
\Lambda dv, \qquad
\theta^a = dx^a.                                      \label{th-qu}
\end{equation}
This defines also the differential of the generators of the algebra
in terms of the basis of the module of 1-forms.  It defines therefore
the conjugacy relations
\begin{equation}
\begin{array}{ll}
[\lambda_+, u] = \Lambda, 
&[\lambda_+, v] = 0,                              \\[4pt]
[\lambda_-, u] = - h \Lambda^{-1}, 
&[\lambda_-, v] = \Lambda^{-1},
\end{array}                                        \label{duality}
\end{equation}
as well as 
$$
[\lambda_a, x^b] = \delta_a^b, \qquad [\lambda_\pm, x^b] = 0 
\qquad [\lambda_a, u] = 0  \qquad [\lambda_a, v] = 0.
$$
From~(\ref{duality}) and the reality conditions it follows that
$$
h^* \Lambda = \Lambda^* h, \qquad \Lambda^* = \Lambda.
$$
Therefore $h^* = h$. In the linearized approximation 
$\Lambda^{\pm 1} = 1 \pm H$.

We shall restrict our consideration to the linear approximation, The
problem is to find an algebra with a set of 4 generators $x^\mu$
satisfying commutation rules as well as a Fourier transformation to a
new set $\lambda_\alpha$ which satisfies Equations~(\ref{consis4}).
We use the map~(\ref{cp}). In the commutative case the geometry is
completely determined (to within integration constants)
by~(\ref{te-te}). We seek an analogous equation in the noncommutative
case. This means that we must rewrite the noncommutative version
of Equation~(\ref{te-te}) in terms of the momenta and compare it with the
Equation~(\ref{s-e2}).  According to the correspondence principle
the form of the $C^\alpha{}_{\beta\gamma}$ as functionals of $h$ are
determined in the commutative limit. A stronger condition is to assume
that they have the same functional form when $\kbar \neq 0$. We
satisfy then automatically the `Bohr-Ehrenfest' condition
\begin{equation}
\lim_{\kbar\to 0} h = \t{h}.
\end{equation}
The non-vanishing structure elements are given by
Equation~(\ref{cp0}) as 
\begin{equation}
e_\alpha h = - C^+{}_{\alpha -}, \qquad
C^\alpha{}_{[\beta\gamma]} + 2i\kbar 
C^\alpha{}_{(\eta\zeta)} P^{\eta\zeta}{}_{\beta\gamma} =
C^\alpha{}_{\beta\gamma}.                              \label{evol1}
\end{equation}
These should be read as equations for $h$ in terms of the commutator
$[\lambda_a,\lambda_+]$ if one believes that the noncommutative
structure determines the gravitational field. If they are read from
right to left then they are equations for the structure of the algebra
in terms of the gravitational field. There is considerable ambiguity
in the Equations~(\ref{evol1}). They are, strictly speaking, only true
in the limit $\kbar\to 0$. We suppose however that this ambiguity can
be absorbed in a `gauge' transformation.

We shall look for a solution with only
$$
H^+_+ = - H^-_- = H \neq 0.
$$
We have then
$$
\theta^+ = du + h dv - H du, \qquad
\theta^- = dv + H dv
$$
as well as
$$
C^+{}_{a-} = -e_a h, \qquad 
C^+{}_{a+} = - C^-{}_{a-} = e_a H.
$$
This in turn implies the relations
\begin{equation}
F^+{}_{a+} = - F^-{}_{a-}, \qquad 
Q^{\pm +}{}_{a+} = - Q^{\pm -}{}_{a-}.                  \label{a-a+}
\end{equation}
To be solved are two sets of equations. The first
is Equation~(\ref{s-e}); for $\alpha \neq -$
\begin{eqnarray}
&&F^+{}_{\alpha -} - 4i\kbar\lambda_r Q^{+r}{}_{\alpha -} 
=  - e_\alpha h, \qquad r = +,-               \label{consis6a}\\[4pt]
&&F^-{}_{\alpha -} - 4i\kbar\lambda_r Q^{-r}{}_{\alpha -} 
=  - e_\alpha H,                               \label{consis6b}
\end{eqnarray}
These are the actual equations; they define the algebra in terms of
the rotation coefficients.  The second set is Equation~(\ref{consis2})
\begin{eqnarray}
&&I_{+-} = K_{+-} + \lambda_r F^r{}_{+-} 
- 2i\kbar \lambda_r  \lambda_s Q^{rs}{}_{+-},       \label{consis6c}\\[4pt]
&&I_{a+} = K_{a+} + \lambda_r F^r{}_{a+} 
- 2i\kbar \lambda_r  \lambda_s Q^{rs}{}_{a+},       \label{consis6d}\\[4pt]
&&I_{a-} = K_{a-} + \lambda_r F^r{}_{a-} 
- 2i\kbar \lambda_r  \lambda_s Q^{rs}{}_{a-},       \label{consis6e}\\[4pt]
&&I_{ab} = K_{ab} + \lambda_r F^r{}_{ab}.           \label{consis6f}
\end{eqnarray}
From Equation~(\ref{consis6a}) we deduce that
$$
e_a e_b h = 4i\kbar I_{ar} Q^{+r}{}_{b-}.
$$ 
Since we are working only to first order we should compare the relative
magnitudes of the various terms in the equations. One can draw the
following table:
$$
\begin{array}{cccccccccc}
[K] &= &L^{-2} &| &[2i\kbar K] &=  &L^0 &\mbox{real} & \\[4pt]
[F] &= &L^{-1} &| &[2i\kbar \lambda F] &=  &L^0 &\mbox{real}
&| &F \sim 2i\kbar\kappa K \\[4pt]
[Q] &= &L^{-2} &| &[(2i\kbar \lambda )^2 Q] &=  &L^0 &\mbox{real} 
&| &Q \sim 2i\kbar\kappa^2 K
\end{array}
$$
We write then
$$
F^r{}_{at} = 2i\kbar\kappa D^{rs}{}_t K_{as},\qquad
Q^{rs}{}_{at} = 2i\kbar\kappa^2 D^{rsu}{}_t K_{au}
$$
with the coefficients $D^{rs}{}_t$ and $D^{rst}{}_u$ real and
without dimension. We have included the factor $2i\kbar$ to indicate
the scaling behaviour and a characteristic gravitational curvature
$\kappa$ to keep the physical dimensions correct and to allow for a
variation in the ratio of the two coefficients. The parameters
$2i\kbar K_{\alpha\beta}$ are real, without dimension and do not
necessarily vanish with the curvature.  We have here implicitly
supposed they do not. This point has been discussed
elsewhere~\cite{MacMadZou02}. From the conditions~(\ref{a-a+}) we see
that the relations
$$
D^{+s}{}_+ = - D^{-s}{}_-, \qquad
D^{+rs}{}_+ = - D^{-rs}{}_-.
$$
We would like the derivative of a physical quantity $f$ to satisfy
$e_a f \sim \kappa f$, which implies that 
$$
(2i\kbar K) \sim \kappa (2i\kbar\lambda).
$$ 
Therefore the order of magnitude of the wave is related to the two
length scales by
$$
h \sim \kappa (2i\kbar \lambda) (2i\kbar K) \sim (2i\kbar K)^2.
$$
If we use the expansion in terms of the coefficients $D^{rs}{}_t$ we
find that
$$
I_{at} = K_{at} - 
2i\kbar\kappa \lambda_r D^{rs}{}_t K_{as} + o(\kappa^2).
$$
To simplify the equations, but without other justification, we choose
$$
K_{+-} = 0, \qquad K_{ab} = 0,  \qquad F^\alpha{}_{ab} = 0,
\qquad F^\alpha{}_{rs} = 0,
\qquad Q^{\alpha\beta}{}_{rs} = 0.
$$ 
The integrability conditions for the equations for $h$ and $H$ are
respectively
\begin{eqnarray*}
&&F^r{}_{ab} e_r h = 4i\kbar I_{[ar} Q^{+r}{}_{b]-},\\[4pt]
&&F^r{}_{ab} e_r H = 4i\kbar I_{[ar} Q^{-r}{}_{b]-}.
\end{eqnarray*}
These can be written as 
$$
D^{rst}{}_- K_{[as} K_{b]t} = 0,\qquad
D^{ps}{}_u D^{uqt}{}_- K_{[as} K_{b]t} = 0
$$
and it implies therefore that
$$
D^{r[st]}{}_- = 0,\qquad D^{p[s}{}_q D^{uqt]}{}_- = 0.
$$
But $D^{rst}{}_-$ must also be symmetric in the first two
indices. We conclude therefore that it is totally symmetric in the
first three indices.

The second set of equations gives rise to Jacobi identities:
$$
\epsilon^{\alpha\beta\gamma\delta}[\lambda_\alpha, I_{\beta\gamma}] = 0.
$$
The Jacobi identities can be written then as
$$
I_{[as} Q^{\pm s}{}_{b]-} = 0.
$$ 
These are the same conditions as for the existence of $h$ and $H$.

We notice also that $h(x^a)$ is harmonic if
\begin{equation}
g^{ab} e_a e_b h = 4i\kbar g^{ab} I_{ar} Q^{+r}{}_{b-} = 0. \label{eq:1}
\end{equation}
This can be written as
$$
g^{ab} D^{+st}{}_- K_{as} K_{bt} = 0,\qquad
g^{ab} D^{ps}{}_q D^{+qt}{}_- K_{as} K_{bt} = 0
$$
To actually find a (formal) solution is more difficult.  One easily
sees that
\begin{eqnarray*}
&&4i\kbar \lambda_r \lambda_s Q^{rs}{}_{a+} = \lambda_+ F^+{}_{a+} -
\lambda_+ e_a H + 4i\kbar \lambda_- \lambda_r Q^{-r}{}_{a+}, \\[4pt]
&&4i\kbar \lambda_r \lambda_s Q^{rs}{}_{a-} = \lambda_r F^r{}_{a-} +
\lambda_+ e_a h + \lambda_- e_a H.
\end{eqnarray*}
Using the first set, the second set of equations can be replaced by
the system
\begin{eqnarray}
&&I_{a+} = K_{a+} + \frac 12 \lambda_+ e_a H
+ \frac 12\lambda_r F^r{}_{a+} - \frac 12 \lambda_- 
(4i\kbar \lambda_r Q^{-r}{}_{a+} - F^-{}_{a+}),     \label{consis7d}\\[4pt]
&&I_{a-} = K_{a-} - \frac 12 (\lambda_+ e_a h 
+ \lambda_- e_a H) + \frac 12\lambda_r F^r{}_{a-}.    \label{consis7e}
\end{eqnarray}
This in turn can be written as
\begin{eqnarray}
&&[\lambda_a, \lambda_+ (1 - \frac 12 H)] = 
(1 - \frac 12 H) K_{a+} +i\kbar\kappa\lambda_r D^{rs}{}_+ K_{as}
\nonumber\\[4pt] 
&&\phantom{[\lambda_a, \lambda_+ (1 - \frac 12 H)]}
- i\kbar\kappa \lambda_- (4i\kbar\kappa \lambda_r 
D^{-rs}{}_{+} - D^{-s}{}_+) K_{as},     \label{consis8d}\\[4pt]
&&[\lambda_a, \lambda_- (1 + \frac 12 H) + \lambda_+ \frac 12 h ] = 
(1 + \frac 12 H) K_{a-} + i\kbar\kappa\lambda_r D^{rs}{}_- K_{as}
\nonumber\\[4pt] 
&&\phantom{[\lambda_a, \lambda_+ (1 + \frac 12 H) + \lambda_+ \frac 12 h ]}
+ \frac 12 h K_{a+}.        \label{consis8e}
\end{eqnarray}
If we suppose that there is a solution $f$ to the equation
$$
e_a f = - F^-{}_{a+} + 4i\kbar\lambda_r Q^{-r}{}_{a+} 
$$
then we can write the second set of equations finally as
\begin{eqnarray}
&&[\lambda_a, \lambda_+ (1 - \frac 12 H) + \lambda_- \frac 12 f ] = 
(1 - \frac 12 H) K_{a+} \nonumber\\[4pt] 
&&\phantom{[\lambda_a, \lambda_+ (1 - \frac 12 H) + \lambda_- \frac 12 f ]}
+ \frac 12 f K_{a-} +i\kbar\kappa\lambda_r D^{rs}{}_+ K_{as},
\label{consis9d}\\[4pt] 
&&[\lambda_a, \lambda_- (1 + \frac 12 H) + \lambda_+ \frac 12 h ] = 
(1 + \frac 12 H) K_{a-} \nonumber\\[4pt] 
&&\phantom{[\lambda_a, \lambda_- (1 + \frac 12 H) + \lambda_+ \frac 12 h ]}
+ \frac 12 h K_{a+} +i\kbar\kappa\lambda_r D^{rs}{}_- K_{as}.
\label{consis9e} 
\end{eqnarray}
Here, as in the previous version we have used the linearization
assumption to replace, for example, on the right-hand side 
$h I_{a+}$ by $h K_{a+}$. By the same assumption we find that
$$
h = -2i\kbar \kappa D^{+s}{}_- \lambda_s,\qquad
H = -2i\kbar \kappa D^{-s}{}_- \lambda_s,\qquad
f = -2i\kbar \kappa D^{-s}{}_+ \lambda_s.
$$
Since there are only $\lambda_\pm$ on the right-hand side, the extra
element $f$ which we have introduced must be a linear combination of
$h$ and $H$. Taking the successive derivatives we find that
\begin{eqnarray*}
&&e_a h = -2i\kbar \kappa D^{+s}{}_- I_{as} \simeq 
-2i\kbar \kappa D^{+s}{}_- K_{as}\\[4pt]
&&e_a e_b h = -2i\kbar \kappa D^{+s}{}_- [\lambda_a, I_{bs}]
\simeq -(2i\kbar \kappa)^2 D^{+s}{}_- D^{rt}{}_s K_{ar} K_{bt}.
\end{eqnarray*}
But we have another expression, Equation~(\ref{consis6b}), for $e_a h$
and the two must be equal. The leading terms are equal by our choice
of normalization. If we compare the next terms, those which dominate
the second derivatives, we find that
$$
2 D^{+rs}{}_- = D^{+t}{}_- D^{rs}{}_t.
$$
In Appendix~I one reads that the metric component $h(\equiv \t{h})$
is a linear function of $u$, harmonic in the $x^a$ and more or less
arbitrary as a function of the $v$.  The term linear in $u$ is in fact
a coordinate artifact and can be eliminated by replacing $u$ and $v$
by new coordinates while maintaining the general metric form. 

Suppose that $h$ has an expansion around a solution of the form of a
vacuum plane wave solution defined by Equation~(\ref{vpws}):
$$
h = h_ax^a + \frac 12 h_{ab} x^a x^b + \cdots
$$
From~(\ref{consis6a}) we find that
$$
F^+{}_{a-} = -h_a, \qquad h_{ab} x^b = 4i\kbar \lambda_r Q^{+r}{}_{a-}
= 2(2i\kbar\kappa)^2 \lambda_r  D^{+rs}{}_- K_{as}.
$$
The relation between the momenta and coordinates is linear as in the
flat-space case. If we take the derivative we find the relations
$$
h_{ab} = 4i\kbar K_{br} Q^{+r}{}_{a-}
= 2(2i\kbar\kappa)^2 D^{+rs}{}_- K_{ar} K_{bs}.
$$
Suppose that the boost $H$ has a similar expansion
$$
H = H_ax^a + \frac 12 H_{ab} x^a x^b + \cdots.
$$
From~(\ref{consis6b}) we find that
$$
F^-{}_{a-} = -H_a, \qquad 
H_{ab} x^b = 4i\kbar \lambda_r Q^{-r}{}_{a-}
= 2(2i\kbar\kappa)^2 \lambda_r  D^{-rs}{}_- K_{as}.
$$
If we take the derivative we find the relations
$$
H_{ab} = 4i\kbar K_{br} Q^{-r}{}_{a-}
= 2(2i\kbar\kappa)^2 D^{-rs}{}_- K_{ar} K_{bs}.
$$
The most interesting relations are those which follow
from~(\ref{duality}). In the present context they become
$$
\begin{array}{ll}
[\lambda_+,u] = 1 + \frac 12 H_{ab}x^a x^b, &
[\lambda_+,v] = 0,\\[4pt]
[\lambda_-,u] = - \frac 12 h_{ab}x^a x^b, &
[\lambda_-,v] = 1 - \frac 12 H_{ab}x^a x^b.
\end{array}
$$
The fact that the right-hand side is quadratic is related to the
approximation we are using and not directly to the quadratic relations
satisfied by the momenta.

When $h=0$, or rather in the limit when $h\to 0$, there is a natural set
of conjugacy relations given by $e_\pm x^a = 0$; $e_a x^\pm = 0$, that
is, with $K_{\pm a} = 0$. When $h \neq 0$ this can no longer be the case.
From the conjugacy relations~(\ref{duality}) we find that
$$
[\lambda_-, x^+] = K_{-a} [x^a, x^+] = -h.
$$
From~(\ref{TE}) below it follows that 
$$
K_{-a} J^{a+} = - h.
$$
Therefore with $h \neq 0$ it is consistent to have $K_{+a} = 0$ as
long as $K_{-a} \neq 0$. We shall examine this possibility in some detail.

We recall first that Equation~(\ref{consis6c})~--~(\ref{consis6f})
should be read from left to right as usual; the $h$ is considered as
been given. We start with the most general possible $h = h(x^\alpha)$.
Only for certain choices of this function is the metric a $pp$-wave
metric. A $pp$-metric arises (by definition) whenever there is a
covariantly constant null vector, for example when the metric is
Ricci flat. To obtain commutation relations for the
position generators one must use the conjugacy
relations~(\ref{duality}). We use Equation~(\ref{pxx}) to obtain
differential equations for $J^{\mu\nu}$:
$$
\begin{array}{ll}
i\kbar dJ^{+-} = [v,h] \theta^-,
&i\kbar dJ^{+b} = [x^b,h] \theta^-, \\[4pt] 
i\kbar dJ^{a-} = 0,
&i\kbar dJ^{ab} = 0.
\end{array}
$$
The Jacobi identities~(\ref{pxx}) reduce then to the equations
$$
e_- J^{+-} = [v,h], \qquad e_- J^{+a} = [x^a,h].
$$
The $J^{\mu\nu}$ must be of the form
$$
J^{+\nu} = J^{+\nu}(u, x^a), \qquad J^{-a} = \theta^{-a}, \qquad
J^{ab} = \theta^{ab}
$$
and they must satisfy the constraint
$$
[v, J^{a+}] = [x^a, J^{-+}]
$$
which follows from the Jacobi identities.  If
Equation~(\ref{J-theta}) is to hold we must have
\begin{equation}
J^{+a} = \theta^{+a} - h \theta^{-a}, \qquad 
h = h(u,v,\lambda_+).\label{TE}
\end{equation}

We have already recalled the fact that in the commutative limit the
commutation relations define a symplectic structure. This is one of
the foundation facts of quantum mechanics.  What is new here is the
fact that the extension of these relations to the differential
calculus leaves also a gravitation field as `shadow'.  We have
constructed the differential calculus (formally) so that it would
define the $pp$ metric in the limit. The two structures have a common
origin and this fact should be apparent.  The $pp$ metric is of Petrov
type $N$. We have found that the matrix~$K_{\alpha\beta}$ - considered
as an electromagnetic field or as a $B$-field - is of type $I$. Both
structures define a common principle null vector, defined by the
propagation of the wave. There remains much to be clarified on the
relation between the two classes of principle null vectors in more
general situations.

We can now consider the algebra and calculus as given and use
previously employed methods to find the metric.

\initiate
\section{Curvature}                                         \label{curv}

The linear-connection coefficients are defined in terms of the
structure elements by the formula~(\ref{T0}), which can be written in
the form
$$
\omega^\alpha{}_{[\beta\gamma]} = C^\alpha{}_{\beta\gamma} - 
i\kbar \omega^\alpha{}_{(\eta\zeta)} Q^{\eta\zeta}{}_{\beta\gamma}.
$$
The case which we shall consider is such however that the second
term vanishes and so we have the usual formula for the Levi-Civita
connection.  The associated covariant derivative is defined by the
actions $D \theta^\alpha = - \omega^\alpha{}_\beta \theta^\beta$ with
the right-hand side defined using~(\ref{t-free}). 

The product in the algebra of forms is not necessarily
anti-commutative but the extra terms will not appear in the expression
for the differential of a form, because of the relation~(\ref{C-con})
and the form of the Ansatz we are using. To calculate the curvature we
must know the structure of the algebra of forms, given by
Equation~(\ref{struc}).  The relations are the usual ones except for
\begin{equation}
\theta^{(+}\theta^{-)} = 4i\kbar Q^{+-}{}_{ra} \theta^r\theta^a ,\qquad
(\theta^+)^2 = 2i\kbar  Q^{++}{}_{ar} \theta^a \theta^r.    \label{acr}
\end{equation}
Because of the symmetries of the rotation coefficients these products
do not contribute to the exterior derivative.

Although not a satisfactory object, we define the curvature exactly
as in the commutative case, as the left-linear map 
$\mbox{Curv} = - D^2$ from $\Omega^1(\c{A})$ into 
$\Omega^2(\c{A}) \otimes \Omega^1(\c{A})$, with associated Ricci map
from $\Omega^1(\c{A})$ into $\Omega^1(\c{A})$ given by
$$
\mbox{Ric} (\theta^+) = -(\Delta h + g^{ab} (e_a h e_b H + e_a H e_b h)) 
\theta^-, \qquad 
\mbox{Ric} (\theta^-) = 0, \qquad
\mbox{Ric} (\theta^a) = 0.
$$
The only
non-vanishing component of the Ricci tensor is therefore
$$
R^+{}_- = -\Delta h - g^{ab} (e_a h e_b H + e_a H e_b h).    
$$ We have not perhaps found the most general $pp$-algebra; in fact we
did not even correctly characterize a $pp$-algebra; but we have shown
there is at least one algebra with a differential calculus whose
compatible metric is a plane wave.

\initiate
\section{Heisenberg  picture}                            \label{Heis}

One could also consider the problem of finding the metric as an evolution
equation in field theory in the sense that one can pass from the
Schr\"odinger picture to the Heisenberg picture with the help of an
evolution hamiltonian. In quantum field theory a scalar field evolves
in time according to the equation
$$
\phi(t) = U(t)^{-1}\phi(0)U(t) = \phi^\prime (0), \qquad U(t) = e^{iHt}.
$$
We have chosen a very particular gauge transformation $\t{\Lambda}$ which
in the commutative limit is one element of the local Lorentz
transformations and we have, in the linear approximation, `lifted'
$\t{\Lambda}$ to a morphism $\Lambda$ of the differential calculus
which is such that the covariant derivative transforms as it
should. We have then on the one hand the transformation of the
rotation coefficients given by Equations~(\ref{evol1}) and on the
other hand what would be the quantum version of a gauge transformation
$$
D\phi \mapsto D^\prime \phi^\prime = 
D^\prime (U^{-1}\phi U) = (U^{-1}D U) (U^{-1}\phi U).
$$
There is a strong similarity between the quantum gauge transformation 
$$
D^\prime = U^{-1}D U
$$
and the noncommutative change of frame~(\ref{evol1}), which supports
the point of view that gravity is a manifestation of the
noncommutative structure of space-time. The evolution of a (free)
quantum field in a gravitational background is the same as, or at
least similar to, a change of frame in a noncommutative geometry.

\initiate
\section{Embeddings}

In this section we would like to consider the $pp$-wave metrics as
perturbations of flat metrics and also consider them as noncommutative
embeddings~\cite{MadMou98,GraLizMarVit01}. To distinguish them we
shall place a prime on perturbed quantities.
The commuting generators can be considered as the coordinates of an
$\b{R}^4$. If $h=0$ one can set $\lambda_a = - K_{a\pm} x^\pm$ to
define an embedding $\b{R}^2 \hookrightarrow \b{R}^4$. Consider now
the 8-dimensional phase-space associated to the $\b{R}^4$ obtained by
adding 4 extra variables $(x^{a}, \lambda_\pm)$ and `quantize' this
space by imposing the commutation relations
$$
[\lambda_\alpha, x^\mu] = i \kbar \delta_\alpha^\mu
$$
We write the corresponding relations Equation~(\ref{duality}) in the
form
$$
[\lambda^\prime_\alpha, x^{\prime\mu}] = i \kbar e_\alpha^{\prime\mu}
$$
and we consider the latter as a perturbation
$$
e_\alpha^{\prime\mu} = \delta_\alpha^\mu + i\kbar f_\alpha^{\mu}
$$
of the former. If we perturb 
$$
x^{\prime\mu} = x^\mu + i\kbar \xi^\mu, \qquad 
\lambda^\prime_\alpha = \lambda_\alpha + i\kbar l_\alpha 
$$
both the position and the momentum generators we see that the relation 
$$
f^{\mu}_\alpha = [\lambda_\alpha, \xi^\mu] + [l_\alpha, x^\mu]
$$
holds. The induced perturbation of the frame is given by~(\ref{lr}) with
$$
\Lambda^\alpha_\beta \simeq \delta^\alpha_\beta - i\kbar f^\alpha_\beta
$$
no longer necessarily a gauge transformation.

If $\lambda_\alpha$ and its perturbed equivalent are to generate
differential calculi then they both must satisfy~(\ref{consis}) with
in general different coefficients. We set
$$
P^{\prime\alpha\beta}{}^{\phantom{\alpha\beta}}_{\gamma\delta} = 
P^{\alpha\beta}{}^{\phantom{\alpha\beta}}_{\gamma\delta} +
i\kbar Q^{\alpha\beta}{}^{\phantom{\alpha\beta}}_{\gamma\delta},
\qquad
F^{\prime\alpha}{}_{\gamma\delta} = F^{\alpha}{}_{\gamma\delta}
 + i\kbar \phi^{\alpha}{}_{\gamma\delta}, \qquad
K^\prime_{\gamma\delta} = K_{\gamma\delta} +i\kbar  \kappa_{\gamma\delta}
$$
and we shall restrict our attention to the special case
$$
P^{\alpha\beta}{}^{\phantom{\alpha\beta}}_{\gamma\delta} =
\frac 12 \delta^{[\alpha}_\gamma \delta^{\beta]}_\delta, \qquad
F^{\alpha}{}_{\gamma\delta} = 0
$$
which is a perturbation of flat space. Comparing then the two sets of
equations we find the relation
\begin{eqnarray*}
&&e_{[\alpha} l_{\beta]} =  \kappa_{\alpha\beta} +
\phi^{\gamma}{}_{\alpha\beta}\lambda_\gamma
- i\kbar l_{[\alpha} l_{\beta]}
- 4 i\kbar Q^{\gamma\delta}{}_{\alpha\beta} 
l_{\gamma} \lambda_{\delta} \\[6pt]
&&\phantom{e_{[\alpha} l_{\beta]}}
-2 \lambda_\gamma\lambda_\delta Q^{\gamma\delta}{}_{\alpha\beta}  
-2 (i\kbar)^2 l_\gamma l_\delta Q^{\gamma\delta}{}_{\alpha\beta}
-2 i\kbar Q^{\gamma\delta}{}_{\alpha\beta} e_{\delta} l_{\gamma} 
\end{eqnarray*}

If the metric is a $pp$-wave then
$$
i\kbar f^\mu_\alpha = \left(\begin{array}{cccc} 
0 &-h &0 &0 \\[4pt] 0 &0 &0 &0 \\[4pt] 
0 &0 &0 &0 \\[4pt] 0 &0 &0 &0 \end{array}\right).
$$
The embedding `space' described by an algebra $\c{B}$ has a natural
differential calculus with a module of 1-forms of rank 8 freely
generated by the elements $dy^i$, with
$$
y^i = (\theta^{\mu\alpha}\lambda_\alpha, x^\nu)
$$
A basis of the 1-forms of the embedded `space' $\c{A}$ would have
four elements $\theta^\alpha$ and would have to be in some sense a
restriction of the 1-forms over $\c{B}$. Associated to the inclusion
of $\c{A}$ in $\c{B}$ there is an inclusion map
$$
\iota^* : \Omega^1(\c{A}) \hookrightarrow \Omega^1(\c{B})
$$
between the respective module of 1-forms. If $\theta^\alpha$ is a
frame and $f$ an element of $\c{A}$ then we must have the relation
$$
0 = \iota^* (f\theta^\alpha - \theta^\alpha f) =
[\iota^* f, \theta^\alpha_i] dy^i.
$$
This implies then the relations 
$$
[\iota^* f, \theta^\alpha_i] = 0.
$$
There are $2n^2$ unknowns and an equal number of relations.

We started with four generators $\lambda_\alpha$ of an algebra $\c{A}$
and an unknown functional $h$ of a second set of four generators
$x^\mu$ of the same algebra. Consider the subalgebra $\c{B}$ generated
by the $x^a$. With the choice we have made for the form of the matrix
$K_{\alpha\beta}$ we can identify $\c{B}$ as a function algebra and
therefore by the commutation relations~(\ref{duality}) we
can conclude that the $\lambda_\pm$ are functions of the
$x^a$. Likewise we identify the $\lambda_a$ as functions of the
$x^\pm$. The four functions constitute an embedding of $\c{A}$ as a
subalgebra of a `flat' algebra $\c{B}$ with twice the
generators. Because of the simple structure of the problem, two of
these functions are just linear transformations.

\initiate
\section{Discussion}

It would seem that it is possible to find a noncommutative geometry,
of Minkowski signature in `dimension' four. We have
started with a given commutative metric, admittedly rather degenerate,
and constructed an algebra and a differential calculus which supports
the metric. The algebra is a `$q$-deformed` version of commutative
$\b{R}^4$; the differential calculus differs from de~Rham's in that the
relations~(\ref{acr}) hold.

Our calculations have been all carried out on the level of formal
algebra. The real problem lies in fact in finding a representation of
the quadratic algebra~(\ref{consis3}). We have given a certain number
of necessary conditions for such a representation to exist; they are
not necessarily sufficient.  To obtain a concrete solution we must
find a representation in terms of operators on a Hilbert space.  To
each representation corresponds a trace and therefore an action.

The Riemann tensor is of null type. Its relation with $J^{\mu\nu}$
would be expressed in terms of the action of $J^{\mu\nu}$ on the
principle null direction. Let $k_\mu$ be this vector. We must compare
$k^\mu$ with $J^{\mu\nu}k_\nu$.  In the limit $\kbar \to 0$ we
recover the commutative plane-wave solution plus in addition a
symplectic structure $J^{\mu\nu}$.  The relation of $J$ to the Riemann
tensor in general is a subject of interest which we leave also to a
future publication.

\appendix

\initiate
\section*{Appendix I}
\setcounter{section}{1}

\renewcommand{\thesection}{\Roman{section}}

For completeness we give here a brief review of plane-fronted metrics
using the 'principal null vector' approach, a formalism which is a bit
more elaborate than is essential but which explicitly identifies, from
the beginning, the important vector $k = \t{e_+}$, with components
$k^{\alpha}$. In special cases $k$ can be identified as a principal
null vector of the Riemann or Weyl (conformal) curvature tensors. We
shall consider $n$-metrics since to do so is no more complicated than
4-metrics ; when $n=4$ these metrics are included in the Kerr-Schild
class of metrics~\cite{KraSteMacHer80}.  We are considering only classical
commutative geometry here so we can omit, without ambiguity, the tilde
notation used in previous discussions of classical metrics. We recall
also that the metric components $g_{ab}$ are real constants.

The line element is given in the frame formalism by
\begin{eqnarray}
ds^{2} = \theta^{-}\otimes\theta^{+} + \theta^{+}\otimes\theta^{-} +
g_{ab}\theta^{a}\otimes\theta^{b},                         \label{ff}
\end{eqnarray}
The class of metrics to be considered will be called 
{\it  plane-fronted}, and can be defined in local coordinates $x^{\alpha}$
by a frame
\begin{equation}
\theta^{\alpha} = (\delta_{\beta}^{\alpha} + 
h k^{\alpha} k_{\beta}) dx^{\beta},
\end{equation}
where the $k^{\alpha}$ are constants given by
\begin{equation}
k^{\alpha}=\delta_{+}^{\alpha},
\end{equation}
so that $\partial_{\beta}(k^{\alpha}) = 0$. Furthermore 
$g_{\alpha\beta}k^{\alpha}k^{\beta} = 0$,  
$g_{\alpha\beta}k^{\beta}=k_{\alpha} =\delta_{\alpha}^{-}$, where, 
in the natural coordinate basis ($x^{+}=u,x^{-}=v,x^{a}$), the flat 
metric takes the form
$$
g_{\alpha\beta}dx^{\alpha}\otimes dx^{\beta} = 2dudv + g_{ab}dx^{a}dx^{b},
$$
and $h$ is a function. Hence the line element~(\ref{ff}) of a {\it
plane-fronted metric} $g$ is given in Kerr-Schild form by
$$
ds^{2} = 2du\otimes dv + g_{ab}dx^{a}\otimes dx^{b} + 2h (dv)^{2}.
$$
Note that the vector field $k^{\alpha}e_{\alpha}=e_{+}$ is a null
vector with respect to these metrics.

The Levi-Civita connection 1-forms, in components with respect to the above
bases ($e_{\alpha},\theta^{\alpha})$ are given by
$$
\omega^{\alpha}{}_{\beta} = (k^{\alpha}\p_\beta h - 
\p^\alpha h k_{\beta})k_{\gamma}dx^{\gamma} =
(k^{\alpha}\p_\beta h -\p^\alpha h k_{\beta})\theta^{-}.
$$
The components of the curvature 2-forms, relative to the same bases,
are given by
$$
R^{\alpha}{}_{\beta\gamma\delta} =
k_{\gamma}k_{\beta}\p_\delta \p^{\alpha}h -
k_{\delta}k_{\beta}\p_{\gamma}\p^{\alpha} h + k^{\alpha}
k_{\delta} \p_{\beta}\p_{\gamma}h - 
k^{\alpha}k_{\gamma} \p_{\beta}\p_{\delta} h.
$$
Consequently
$$
R^{\alpha}{}_{\beta\gamma\delta} k^{\delta} = 
k_{\gamma}k_{\beta}\p_\delta \p^{\alpha} h k^{\delta} - 
k^{\alpha}k_{\gamma}\p_{\beta}\p_{\delta} h k^{\delta},
$$
and the Ricci tensor has components
$$
 R_{\beta\delta} = R^{\alpha}{}_{\beta\alpha\delta}
= k_{\alpha}k_{\beta}\p_\delta \p^{\alpha} h - 
k_{\delta} k_{\beta}\p_\alpha \p^{\alpha} h + 
k^{\alpha} k_{\delta} \p_{\beta}\p_{\alpha} h.
$$
Hence
$$
R_{\beta\delta}k^{\delta}=0 \quad \mbox{iff} \quad
\p_{\beta}\p_{\delta}hk^{\delta}k^{\beta}=0.
$$
The results above can be used to prove the following theorems which
illustrate some of the properties of the class of metrics being
considered here~\cite{KraSteMacHer80,MisThoWhe73,EhlKun62,Pir65,Gri91}.

\begin{theorem}
  
Let $D_\alpha$denote the Levi-Civita covariant derivative
(components with respect to the moving frame). Then

\begin{enumerate}

\item The following equality holds

\begin{equation}
D_\alpha k_{\beta}=k_{\alpha}k_{\beta}\p_{\gamma} h k^{\gamma}.
\end{equation}

\item $k^{\alpha}D_{a}k_{\beta}=0,$ and $k^{\alpha}e_{\alpha}$ is
  tangent to affinely parametrized null geodesics.

\item $D_{[\alpha}k_{\beta]} = D_{\alpha}k^{\alpha}=0$; \quad 
$D_{\alpha}k_{\beta} = D_{(\alpha}k_{\beta)}$ \quad \hbox{and} \quad 
$D_{\alpha}k_{\beta} D_{}^{\alpha}k^{\beta} = 0$. 

\end{enumerate}
\end{theorem}

This justifies the teminology `plane-fronted', since these conditions
applied to vacuum Lorentzian 4-metrics of the form considered here,
define the so-called {\em plane-fronted metrics}

\begin{theorem}

$D_{a}k_{\beta}=0$ if and only if
\begin{equation}
\p_+ h = \p_\beta h k^{\beta}=0.
\end{equation}

\end{theorem}

Once again using the previously mentioned Lorentzian 4-metric
terminology, when this equation is satisfied, the plane fronted
metrics which admit a covariantly constant null vector
field are said to be {\em plane-fronted with parallel rays
($pp$~waves)}.

\begin{theorem}

Consider {\em plane-fronted} metrics. Two useful results are the following.

\begin{enumerate}

\item $\ R^{\alpha}{}_{\beta\gamma\delta}k^{\delta}=0$ if and only if%
\begin{equation}
\p_{\alpha}\p_{\beta} h k^{\beta} = Fk_{\alpha},
\end{equation}
for some function $F$, and then the Ricci tensor must take the form%
\begin{equation}
 R_{\alpha\beta} = k_{\alpha}k_{\beta}(2F-\p_\alpha \p^{\alpha} h).
\end{equation}

\item  Let $Q_{\alpha\beta}=k_{\alpha}q_{\beta} - k_{\beta}q_{\alpha}$ 
where $k^{\alpha}q_{\alpha}=0$ and the $q_{\alpha}$
are also constants. That is, $\p_{\beta} q_{\alpha} = 0$. 
Then $Q_{\alpha\beta}k^{\beta} =0$ and, if $f$\ is any function,
\begin{equation}
D_\alpha (fQ_{\beta\gamma}) = Q_{\beta\gamma}(\p_{\alpha}f - 
hk_{\alpha} \p_{\beta}f k^{\beta} +
fk_{\alpha}\p_\beta h k^{\beta}).
\end{equation}

Hence when $fQ_{\beta\gamma}$ is non-zero $D_\alpha(fQ_{\beta\gamma})=0$ if
and only if the equation%
\begin{equation}
\p_\alpha \log |f| + k_{\alpha}\p_\beta h k^{\beta}=0,
\end{equation}
admits a non zero solution $f$.
\end{enumerate}

\end{theorem}
By using the results above it is a straight-forward matter to obtain further
results about particular classes of solutions of Einstein's equations (see
further comment below).

\begin{theorem}                                         \label{th4}

\begin{enumerate}

\item In any dimension greater than two, plane-fronted metrics are Ricci
flat, $R_{\alpha\beta}=0,$ if and only if the two equations%
$$
\p_{\alpha}\p_{\beta}k^{\beta} h  = 
F k_{\alpha}, \qquad \p_\alpha \p^{\alpha} h = 2F
$$
for some function $F$, are satisfied.

Hence it follows from the above results that when the latter two
equations are satisfied, the Weyl conformal tensor satisfies the
equation $C_{\beta\gamma\delta}^{\alpha} k^{\delta}=0$ and, in the
four dimensional Lorentzian case, is said to be of Petrov type $N$ with
$k^{\alpha}e_{\alpha}$ being a repeated principal null vector.

\item Using the local coordinate expressions further it follows
  immediately that {plane-fronted metrics are Ricci flat} if and only
  if the function $h$ is given by%
\begin{equation}
h = L(v) u + G(v,x^{a}),
\end{equation}
where $L(v)$ satisfies $\p_v L = F$ and the function  $G$
satisfies the (Laplace-type) equation
\begin{equation}
g^{ab} \p_{a}\p_{b} G = 0.                          \label{laplace}
\end{equation}

\item In the {four dimensional Lorentzian case}, writing $z=x+iy,$ it
follows from these equations that plane fronted metrics g is 
{\it Ricci flat} if and only if
\begin{eqnarray}
G = K(v,z)+c.c., \qquad h = Lu+G(v,x^{a}).        \label{dcr}
\end{eqnarray}
for {any} complex $K$ holomorphic in z.

\end{enumerate}
\end{theorem}

It is a straightforward matter to show that the term $Lu$ in (8.10)
and (8.12) is merely an artifact of the coordinate choice. It can be
eliminated, while retaining the general form of the metric, by
choosing new $u,v$ coordinates.  Hence the Ricci-flat solutions are
$pp$-waves. As an immediate corollary follows the

\begin{theorem} 

In the four-dimensional Lorentzian case, using the notation of
Theorem~\ref{th4}, it therefore follows that {\em Ricci flat $pp$ waves} are
determined by an arbitrary complex function K where
\begin{equation}
h = G(v,x^{a}) = K(v,z) + c.c.
\end{equation}

\end{theorem}

In summary

\begin{enumerate}

\item Functions $h=h(v)$ give flat metrics as do functions linear in the
coordinates $x^{\alpha}$.

\item It is clear from the form of the Ricci tensor given above that
  is is a straightforward matter to produce plane-fronted (or more
  particularly $pp$-wave) metrics which are solutions of the Einstein
  field equations with matter sources such as a massless scalar field
  ($R_{\alpha\beta}=\kappa e_{\alpha}(\varphi )e_{\beta}(\varphi),$
  with say $\varphi=\varphi(v)$ ) and gauge fields (for example the
  null electromagnetic field with components
  $F_{\alpha\beta}=k_{\alpha} p_{\beta}-k_{\beta}p_{\alpha}$, with
  $k_{\alpha}p^{\alpha}=0,$ so that $F_{\alpha\beta}F^{\alpha\beta}=0$
  and the energy momentum tensor is (modulo sign conventions) given by
  $k_{\alpha}k_{\beta} (p_{\gamma}p^{\gamma})$).
  
\item One final specialization made in the literature which is
  relevant here is from $pp$ waves to {\it plane waves.}. This
  corresponds to making particular choices of solutions of the
  equations above. In the vacuum case , the solutions of
  Equation~(\ref{laplace}) given by
\begin{equation}
h = \frac 12 h_{ab}(v)x^{a}x^{b}, \qquad g^{ab} h_{ab} = 0, \label{vpws}
\end{equation}
are {\it vacuum plane wave solutions}. This terminology can be
justified by computing the Weyl tensor and noting the planar symmetry.
In addition, the notions of amplitude and polarisation can be
introduced.

\end{enumerate}

\initiate
\section*{Appendix II}
\setcounter{section}{2}

An alternative form of a plane wave metric, derivable as the Penrose
limit of any 4-metric, is currently of some interest in the
string-theory community~\cite{BlaFigHulPap02}. It can be obtained by
the following argument.  Consider an arbitrary metric in null
coordinates $x^\mu = (u,v,x^i)$. The hypersurfaces $u = u_0$ are
assumed to be null and $v$ to be the affine parameter along the
corresponding bicharacteristics.  The line element can be written in
the form
$$
ds^2 = 2 du dv + g_{++} du^2 + 
2 g_{+i} dv dx^i + g_{ij} dx^i dx^j.
$$
Since $(0,1,0,0)$ is a null vector we have $g_{--} = 0$. We have
also used the local coordinate freedom to set
$$
g_{+-} = 1, \qquad g_{-i} = 0.
$$
The remaining metric components are general functions of all
coordinates. We now rescale
$$
g_{\mu\nu} \mapsto \omega^{-2} g_{\mu\nu}, \qquad
u \mapsto \omega^2 u, \qquad x^i \mapsto \omega x^i.
$$
The line element becomes
$$
ds^2 = 2 du dv + \omega^2 g_{++} du^2 + 
2 \omega g_{+i} du dx^i + g_{ij} dx^i dx^j
$$
with the metric components now functions of the form
$$
g_{\mu\nu} = g_{\mu\nu}(\omega^2 u, v, \omega x^i).
$$
As a singular limit when $\omega \to 0$ one
obtains~\cite{Pen76,BlaFigHulPap02} the plane wave with line element 
$$
ds^2 = 2 du dv + g_{ij}(v) dx^i dx^j.
$$
If the initial metric is described by a frame of the form
$$
\theta^+ = du, \qquad
\theta^- = dv + \theta^-_+ du +  \theta^-_i dx^i,\qquad 
\theta^a =  \theta^a_i dx^i
$$
then the scaled metric is described by
\begin{equation}
\theta^+ = du, \qquad
\theta^- = dv, \qquad \theta^a =  
\theta^a_i(v) dx^i.                         \label{cR-frame}
\end{equation}
We shall refer to~(\ref{cR-frame}) as the Rosen frame. 

In the `quasi-commutative' limit the line element is given by
$$
ds^2 = 2d\t{u} d\t{v} + 2\t{h} d\t{v}^2 - d\t{x}^2 - d\t{y}^2.
$$
In the Rosen frame it is of the form
$$
ds^2 = 2du dv - (\theta^1)^2 - (\theta^2)^2.
$$
This form of the metric has a certain similarity to the Kasner
metric, which has also~\cite{MacMadZou02} a noncommutative extension.

The dual frame is such that the only nontrivial commutation
relations are given by
\begin{equation}
[e_-, e_a] =   (e^{-1} \dot e)^b{}_a e_b, \qquad 
\theta^a_i e^i_b = \delta^a_b.                                   \label{crs}
\end{equation}
The structure functions $C^\alpha{}_{\beta\gamma}$ are seen to be with
$$
C^a{}_{-b} = - C^a{}_{b-} = (e^{-1} \dot e)^b{}_a. 
$$
the only non-vanishing components.

The transformation from the Rosen frame to the previous frame for the
tilded metric is made using the transformation of coordinates
$$
u = \t{u} +\frac 12 F_{ab} (\t{v}) \t{x}^a \t{x}^b, \qquad
v = \t{v}, \qquad
x^i = e^i_a (\t{v}) \t{x}^a.
$$
In the old coordinates the Rosen frame is of the form 
$\theta^\alpha = \Lambda^\alpha{}_\beta \t{\theta}^\beta$ with
$$
\Lambda^\alpha{}_\beta = \left(
\begin{array}{ccc}
1 &0 &-F_{bc}\t{x}^c \\[2pt]
0 &1 &0 \\[2pt]
0 &(e^{-1}\dot e)^a{}_c \t{x}^c &1
\end{array}\right)
$$
The matrix $\Lambda^\alpha{}_\beta$ is a local Lorenz transformation
if
$$
g^{+a} = \Lambda^+{}_\alpha \Lambda^a{}_\beta g^{\alpha\beta} =
\Lambda^a{}_- + \Lambda^+{}_b g^{ab} = 
(e^{-1} \dot e - F)^a{}_b \t{x}^b = 0.
$$
That is, if 
$$
F =  e^{-1} \dot e
$$
Here the dot denotes derivative with respect to $v$.  It should be
noted that, because of the special form of $h$ one obtains, this
applies to the special case of plane waves.

The matrix $F^a{}_b(v)$ is an arbitrary real element of the
algebra $M_2$ of complex $2 \times 2$ matrices, which we write
$$
F = S + A
$$
as the sum of a symmetric part and an antisymmetric part.  The
latter defines a rotation in the plane transverse to the null
coordinates.  The non-vanishing components of the connection form are
given by
$$
\omega^a{}_- = -S^a{}_b \theta^b,  \qquad
\omega^+{}_a = S_{ab} \theta^b,  \qquad
\omega^a{}_b = A^a{}_b \theta^-.
$$
The curvature 2-form has 
$$
\Omega^a{}_- = - (\dot S^a{}_b + [A, S]^a{}_b - (S^2)^a{}_b)  
\theta^- \theta^b, \qquad \Omega^a{}_b = 0
$$
as non-vanishing components. The non-vanishing components of the
Riemann tensor are given by
$$
R^a{}_{--b} =  - (\dot S + [A, S] - S^2)^a{}_b
$$
as well as those obtained from symmetries. The only non-vanishing
component of the Ricci tensor is
$$
R_{--} = - \tr (\dot  S + [A, S] - S^2).
$$
So the field equations reduce to the condition that the trace of
a matrix be constant.  The solution has three arbitrary functions; two
dynamical ones in $S$ and a gauge degree of freedom in $A$.

\initiate
\section*{Acknowledgments} This work was to a large extend carried out
while one of the authors (J.M.) was visiting the MPI in M\"unchen. He
is grateful to Julius Wess for financial support during this period.
D.C.R. would like to thank the Theoretical Physics Group of Imperial
College, London for its hospitality during this time. The authors
would like to thank Lane Hughston for discussions.


\providecommand{\href}[2]{#2}\begingroup\raggedright\endgroup

\end{document}